\documentclass{aastex631}
\usepackage{todonotes}
\usepackage{booktabs}
\usepackage{longtable}
\usepackage{array}
\usepackage{amsmath}
\usepackage{cleveref}

\usepackage{epsfig}
\usepackage{graphicx}
\usepackage{graphicx,color}
\pdfoutput=1
\def\simpropto{\lower.2ex\hbox{$\; \buildrel \propto \over \sim \;$}}
\def\ltsim{\lower.5ex\hbox{$\; \buildrel < \over \sim \;$}}
\def\gtsim{\lower.5ex\hbox{$\; \buildrel > \over \sim \;$}}

\begin{document}
%\pagestyle{myheadings}
%\thispagestyle{empty}
%\normalsize

\title{Which came first: supermassive black holes or galaxies? Insights from JWST}

\correspondingauthor{Joseph Silk}
\email{silk@iap.fr}

\author{Joseph Silk}
\affiliation{Institut d'Astrophysique, UMR 7095 CNRS, Sorbonne Universit\'{e}, 98bis Bd Arago, 75014 Paris, France} 
\affiliation{Department of Physics and Astronomy, The Johns Hopkins University, Baltimore MD 21218, USA}
\affiliation{Beecroft Institute of Particle Astrophysics and Cosmology, Department of Physics, University of Oxford, Oxford OX1 3RH, UK}
\author{Mitchell C.~Begelman}
\affiliation{JILA, University of Colorado and National Institute of Standards and Technology, 440 UCB, Boulder, CO 80309-0440, USA} 
\affiliation{Department of Astrophysical and Planetary Sciences, 391 UCB, Boulder, CO 80309-0391, USA}
\author{Colin Norman}
\affiliation{Department of Physics and Astronomy, The Johns Hopkins University, Baltimore MD 21218, USA}
\author{Adi Nusser}
\affiliation{Department of Physics and Asher Space Research Institute, Israel Institute of Technology, Technion, Haifa 32000, Israel}
\author{Rosemary F.~G.~Wyse}
\affiliation{Department of Physics and Astronomy, The Johns Hopkins University, Baltimore MD 21218, USA}
 
%keywords: XXXXX
%\author{Rosemary F.G.~Wyse}
%\affil{Department of Physics and Astronomy, The Johns Hopkins University}
\date {\today}

%\maketitle  %OMIT IN 2 COL VERSION
\begin{abstract}
%Supermassive black hole  formation may have preceded, coevolved with, or followed  the birth of star forming galaxies.  

Insights from JWST observations suggest
%are shedding new light on the chronology and nature of AGN in the context  on galaxy evolution. We argue 
that AGN feedback evolved from a short-lived, high redshift phase in which  radiatively cooled turbulence and/or momentum-conserving outflows 
%in  dusty ultracompact galaxy hosts. 
%These relatively dense outflows 
stimulated vigorous early star formation (``positive'' feedback), to late, energy-conserving outflows that depleted  halo gas reservoirs and quenched star formation.  The transition  between these two regimes occurred at $z\sim 6$, independently of galaxy mass, for simple assumptions about the outflows and star formation process.   Observational predictions provide circumstantial evidence for the prevalence of massive black holes at the highest redshifts hitherto observed, and we discuss their origins.
%, with seeds requiring a scale-dependent  boost in the power spectrum of the primordial density fluctuations on 
%\mitch{What does ``boosted" mean here?  Do you mean that the these accretion ionto these fluctuations is boosted or something else?  The word seems ambiguous.}
%small scales,  or possibly the presence of rare primordial black holes of intermediate mass.
        %, complementing  constraints from Eddington-limited growth of SMBHs,  for  possibly primordial and subdominant intermediate mass black hole seeds.
% %only contributing  to $\sim10^{-4}$  of dark matter  in the observationally constrained $10^4-10^6\rm M_\odot$ mass range. 
\end{abstract}

\keywords{galaxies:formation; galaxies:nuclei; quasars:general; galaxies:jets; stars:formation }

\section{Introduction}

The first year of JWST science is changing our perspective on high-redshift galaxy formation. There is an intimate connection with AGN that is not yet effectively incorporated into current simulations.  Broad-line AGN seem ubiquitous at high redshift and their host galaxies are often ultracompact and dust-reddened \citep{greene23}.  
%The spectroscopically  confirmed AGN have black hole masses of $10^7 -10^9\;  \rm M_\odot $  and inferred abundance two orders of magnitude higher than that of the faintest UV-selected quasars and some 1\% of the UV-selected
%magnitude higher  than the abundance of 
%galaxies at $z\gtsim 5$ \citep{matthee23}. 
 
The spectroscopically  confirmed AGN have black hole masses of $10^7 -10^9\;  \rm M_\odot $  and are inferred to be two orders of magnitude more abundant than the faintest UV-selected quasars, yet they represent only 1\% of the UV-selected (star-forming) 
galaxies at $z\gtsim 5$ \citep{matthee23}.
Their compactness is key: we argue that cooling times in the AGN-shocked gas are significantly reduced, leading to
shock-boosted star formation by compression of ambient gas clouds via mechanical feedback from  fast outflows. 

At present no one dataset for high-redshift galaxies constrains all the important parameters that underlie the unified theory we develop below. Different samples provide complementary insights so for present purposes we assume broader applicability of observed behavior. This will allow us to make predictions to test whether our assumptions are valid. 

We note that galaxies at high redshift with sufficient cooling to be in the 
positive feedback stage all have their central AGN obscured. Since 
 they are $\mathcal O{(0.1-1)} $   Compton thick, we will argue that 
new ways of finding them directly (e.g.~SKA, future very sensitive X-ray missions) or indirectly (e.g.~very high %triggered 
specific star-formation rates [sSFR]) are needed.
 
Star formation rates are enhanced and exceptionally luminous  galaxies (compared to current models) are found at $z\gtsim 10 $ \citep{harikane23}. % Another intriguing result is that 
%The local high redshift environment appears to be metal-poor. 
Yet the stellar content of massive galaxy halos seems to be depleted when the black hole masses are compared to stellar and dynamical masses, respectively \citep{maiolino23}.  We infer that most of the stars have yet to form, but
that the normal amount of gas is present in standard halos.  We further infer that AGN luminosities most likely dominate in the ultraluminous galaxies, at least in those offset from the  $M_{BH}-M_\ast$ scaling relation at redshifts $z\gtsim 5$.

Chemical signatures in the highest redshift galaxies \citep{bunker23} mimic those found in quasar emission line regions \citep{dietrich03} as well as in some massive star clusters. These observational  results, while still based on sparse data and subject to large uncertainties \citep{wang23}, motivate a fresh look at the coevolution of galaxies and supermassive black holes (SMBHs). They suggest an intimate connection between black hole feedback and rapid (massive) star formation, indicative of 
what is often referred to as positive feedback, because the chemical timescales must be relatively short.
%, without enough time for AGB star formation. 
%\mitch{I'm still not sure about this last statement. Surely, massive stars would go into the AGB stage almost instantaneously (few Myr) compared to the other timescales we're discussing.}  
This connection, resulting in the speed-up of chemical evolution, would  be further  enhanced by including the effects of stellar tidal disruptions as suggested to account for the [N/C] enhancement and abundance pattern consistent with a single TDE event in a nearby AGN \citep{miller23}, or  by the possible role of very massive or even supermassive stars \citep{vink23, marques23}.
 
Let us first discuss how feedback might be positive or negative.  The idea is  that in  a star-forming galaxy that hosts an AGN 
\begin{itemize}
    \item 
    the accumulation of gas in the central region initially enhances both accretion onto the central SMBH and star formation.
    
    \item  black hole outflows are invigorated, driving shocks and turbulence into nearby gas.
    
    \item  cooling of shocked gas is effective, leading to a dense, cool phase where star formation is boosted. Feedback is positive.
    
    \item  at late times, cooling is ineffective, energy-conserving winds drive gas outflows, and feedback on star formation is negative as the gas reservoir is depleted.
\end{itemize}

We argue that the high redshift SMBH-containing galaxy population is ultracompact (Sec.~2), leading to rapid cooling of outflow-shocked gas, triggering  star formation through positive feedback. This transitions to negative feedback at lower $z$, as cooling becomes inefficient and star formation is quenched by energetic outflows. In Sec.~3 we estimate the transition redshift and describe the resulting SMBH scaling laws and observational implications. The sequence of events we propose, and their approximate timing, are illustrated in Fig.~1.  We discuss possible early sources of massive black hole seeds in Sec.~4 and  summarize our results in Sec.~5.
\begin{figure*}
    \includegraphics[width=\textwidth]{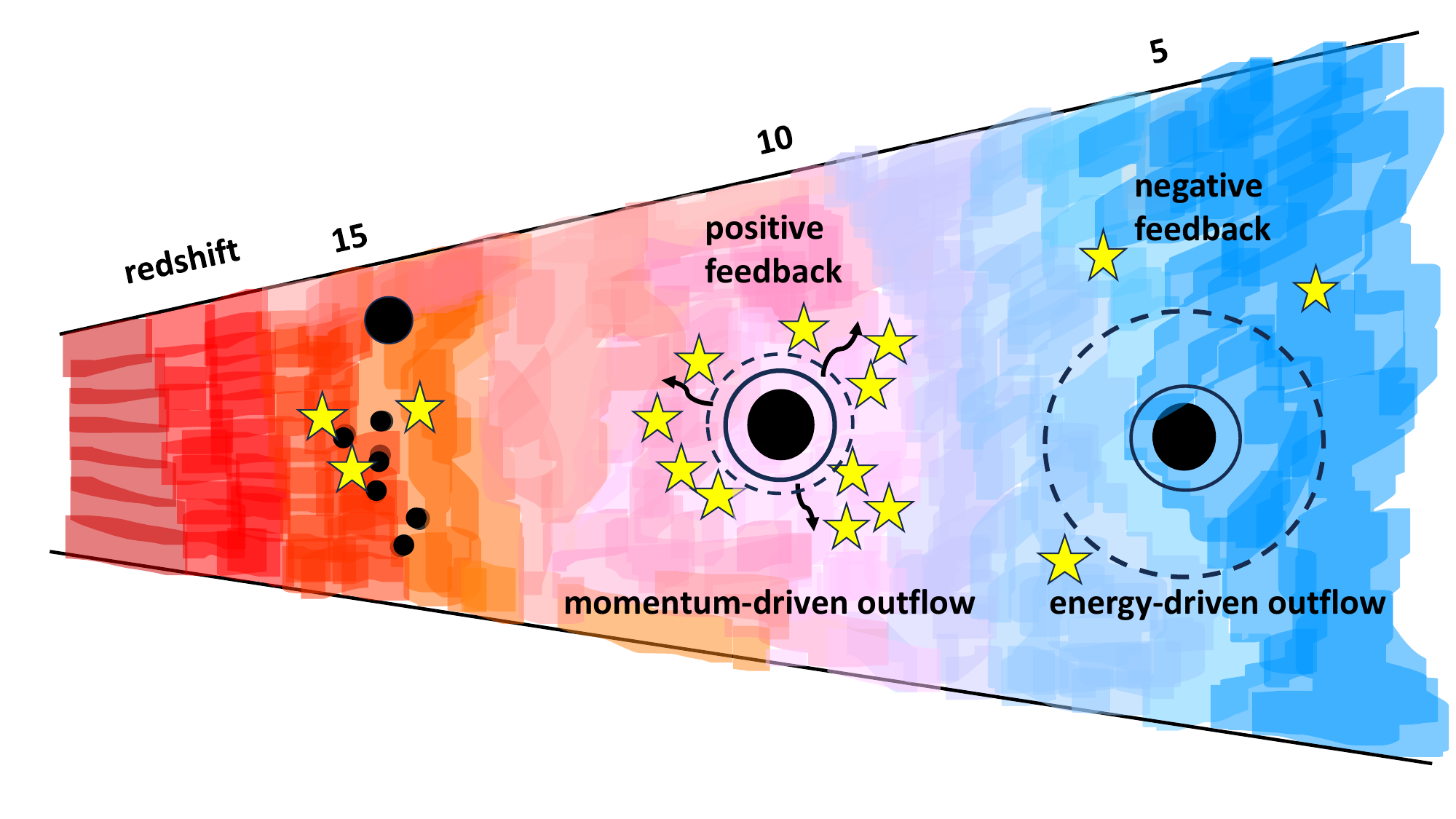}
   % {%figures/retII_allthree_ks_smallpop_1.jpeg}
    \caption{The transition in star formation rates and black hole growth as redshift decreases from regimes where positive feedback dominates to a later epoch when feedback is largely negative.  
    (After Fig.~2 of \citealt{costa14}.)  The epoch of reionization is indicated by the change from red to blue. }
    \label{fig:cartoon}
\end{figure*}
\\ \\ \\

\section{Compactness drives cooling and triggering of star formation} 

The newly discovered population of ultracompact red galaxies --- ``little red dots'' --- at 
$10\gtsim z\gtsim 5$ from JWST data exhibits effective radii of only around $150 \; \rm pc$ yet are the natural candidates for massive galaxy precursors, with central stellar densities comparable to or exceeding  those of nearby ellipticals \citep{baggen23}.  These are lower mass galaxies 
 $\sim 10^9-10^{10} \rm M_\odot $, whereas resolution of stellar light in massive quasar host galaxies at 
 $z>6$ \citep{ding23} suggests that these systems have radii of $\sim 1$ kpc, still
 significantly smaller than the several kpc scales of nearby massive ellipticals.
 
%\AN{\textbf{A caveat:} Since the stellar density is the same in high z galaxies and near ellipticals, may imply that there nothing special  about the SF at high redshift and that it is the same as low z but happened early. } 
The best size/mass measurement for a little red dot comes from \cite{furtak23} for the triply lensed A2744-QSO1 source, only 30 pc in size and  with a quoted stellar mass of $\ltsim 2\times 10^9 \rm M_\odot$ and spectroscopically measured SMBH mass of $3\times 10^7 \rm M_\odot$ at $z\sim 7$.  The puzzle is how such high stellar densities and large black hole masses, relative to stellar mass, are present at such high redshift.  We address this in Section 4.

Many little red dots are dust-reddened AGN with black hole masses in the range $10^7 -10^8 \; \rm M_\odot$ \citep{greene23}.  We note that, if we plausibly conflate these with compact red galaxies, \cite{baggen23} state that 4/9 of such objects could well host AGN, but this is indeed  an open question.  We predict they should contain AGN. 
The high number density, some 100 times higher than the faintest UV-selected quasars, and near-Eddington luminosities, suggest that they may play a key role in black hole evolution at high redshift,  as we now argue.  These estimated numbers of high redshift AGN may be significantly underestimated because of dust obscuration, especially in ultracompact host galaxies \citep{andonie23}.

Mechanical feedback can be driven in both AGN modes: in {\it radio mode}, through powerful, fast jets \citep{cielo18}; and in {\it quasar mode} \citep{giustini19}, through disk winds which may be radiatively driven.  Generic mechanical feedback arises from both modes and energizes the surrounding gas, which either cools or does not.  When cooling is efficient, energy injection may take the form of widespread virial turbulence, balanced by local radiative losses, or --- if the ambient medium is sufficiently diffuse --- momentum-conserving radiative shocks.  When cooling is inefficient, feedback leads to a fast, energy-conserving flow that escapes the central region of the galaxy. 
%In fact the Compton temperature is of order the virial temperature. 
%The dominance of mechanical input by driving fast shocks may  play a key role in developing momentum-conserving outflows if bremsstrahlung cooling is dominant in the circumnuclear regions, where the inferred dense interstellar gas is likely to be Compton-thick \citep{gilli22}.
%{ This so-called {\it radio} mode may be important for the most luminous AGN with strong jets  in clumpy media \citep{cielo18}, and similar conclusions apply to disk-driven  winds from radio-quiet quasars \citep{giustini19}.}
%\mitch{As far as I can tell the Cielo paper is just wrong, since it neglects what we know about disk winds from radio-quiet quasars.  BALs studies show that all radio-quiet quasars have these winds, and thet they are probably powerful. The Cielo model has nothing to do with Compton heating and they seem to leave out radiation pressure on strong resonance lines, which is the main effect producing the feedback. As for radio mode, I suspect that this form of fewedback is less important than disk winds from radio-quiet quasars, both because of their directionality and their relative rarity. BUt I think this distinction is irrelevant to our argument --- all that matters is the cooling (or not) of the stirred up ISM.}

The feedback can terminate black hole growth by accretion at sufficiently large black hole masses, and also regulate or quench growth of the galactic stellar mass,  leading to a universal scaling relation between black hole mass and stellar velocity dispersion.  Whether the black hole can be said to regulate galaxy growth, or vice-versa, depends on initial conditions as well as the details of cooling efficiencies and outflow time scales.  It simplifies matters to consider two limiting cases, depending on the role of cooling as the AGN outflow propagates through the surrounding interstellar medium.  The first case features rapid cooling compared to the crossing time of turbulent motions in the ultracompact protobulge, and characterizes early times; in the second case, valid at later times, cooling is ineffective in the feedback-expanded, now conventionally-sized, gas-rich bulge.  We assume that the ultracompactness at high redshift evolves, via merging and accretion, into relatively normal early-type galaxies  at low redshift, perhaps terminating their growth by cosmic noon at $z\sim 2$.   

Let us apply this comparison of dynamical and cooling timescales to the protobulge, which we assume is of order the median   observed radius $r_{150} =r_{\rm eff} /150$ pc for a galaxy mass normalized to $M_{10} = \rm M/10^{10} M_\odot$. First, we summarize the various parameters: 

\begin{itemize}

  \item 
  The mean particle density is  $\bar n \approx 3M/4\pi \mu m_p r^3 = 6 \times 10^4 \, M_{10}r_{150}^{-3} \rm \, cm^{-3}$, corresponding to a column density  $N \approx \bar n r = 3 \times 10^{25} \, M_{10}r_{150}^{-2} \rm \, cm^{-2}$. 
  
  \item 
  The virial temperature is $T \approx GM\mu m_p/2 kr =  10^{7} \,M_{10} r_{150}^{-1} \, \rm K $. 
  \item The dynamical timescale  is $t_{\rm dyn} \approx (r^3/GM)^{1/2 }= 3 \times 10^5 \, r_{150}^{3/2}M_{10}^{-1/2} \, %\delta_c^{-1/2}
  \rm yr$. 
  %\mitch{$kT/\mu m_p = \sigma^2$ then gives $10^{13}$ s  $= 3\times 10^5$ yr. Should use yr for both dynamical and bremss cooling time. }
  
  \item 
  The bremsstrahlung cooling rate is 
$ \lambda_{\rm ff}\, n^2 T^{1/2} \rm erg \ cm^{-3}
 s^{-1}$ where $\lambda_{\rm ff}=1.4 \times 10^{-27} $ in cgs units and the corresponding cooling timescale is $t_{\rm cool}= 5 kT^{1/2} /2 \mu\lambda_{\rm ff}n  \approx 300 \, r_{150}^{5/2}M_{10}^{-1/2} \,\rm yr$.
 
  \item
  The post-shock column density of cooled gas behind shocks at close to the virial speed, $v_s \approx  600 \, (M_{10}/ r_{150})^{1/2}$ km s$^{-1}$, is at least $N_{\rm cool} = n t_{\rm cool} v_s \approx 2 \times 10^{22} \, M_{10} r_{150}^{-1}$ cm$^{-2}$.
  
 \end{itemize}

Traversing this critical column density is key to our argument below on the transition between the momentum and energy-conserving regimes.  Note that in the likely presence of clouds, the cooling timescale is reduced by the density contrast, which could further reduce the local cooling timescale by an order of magnitude or more. In general, we are in a cooling-dominated regime, especially for the ultracompact galaxies, arguing strongly for radiatively damped turbulence or momentum-driven flows in the gas-rich systems we are considering at high redshift.
 
%\AN{notation needs to be consistent with the previous paragraph. There really is room for improvement in this central segment of the text. Why $t_{br}=10^2 n_6^{-1} \rm yr,$ here while if you substitute $n\sim 10^{6}/cm^3$in the expression $t_{br}=10^{13}M_{10}^{1/2}n_4^{-1} r_{150}^{-1/2} \rm s$  (above) you get $10^{11} \rm s\approx 3000yr$}
%{\color{red} The post-shock temperature for an outflow at $\sim 0.1 c$    is $\sim 10^{11}\rm K.$ The density in the gas-rich precursor is $\sim 10^6 {\rm cm}^{-3} M_{10}r_{150}^{-3}$  and the post-shock cooling time is $t_{br}=10^2 n_6^{-1} \rm yr,$ to be compared with the local dynamical time  of a typical "red dot" galaxy $t_{dyn} \sim 3. 10^{6} n_6^{-1/2}\rm yr.$ Here $n_6 =n/10^6\rm cm^{-3}$  and $r_{150} =\rm r/ 150 pc.$ }
%\AN{I must admit I am confused here. in a preceding paragraphs, a $10^7K$ is mentioned for the shocked gas.Must sharpen  the discussion of the various stages of the evolution. }
%I am also confused about the significance of the radius $R_c$. It seems to me that the relevant radius is  $R_1$ in paper [8] since it is the radius where $T>T_v$. }

\section{Transitioning from high to low $z$}
%Cosmological time-scales (time available for galaxies to form) $t_c$ are $1.2$ Gyr at $z=5$, $0.5$ Gyr at $z=10$, and only $0.3$ Gyr at $z=15$. The latter corresponds to about 10 times the main sequence age of a $10 \; \rm M_\odot$ star. Assuming that it takes of order a few protogalactic rotation times  to  shed  initial angular momentum and form the first stars, we infer this is a  rough lower limit to stellar masses in the first generation of galaxies at $z\sim 15$. \mitch{I don't understand this argument.  Why does the MS lifetime care about the formation time.  The Kelvin contraction time for a 1 solar mass protostar is only a few million years so they can effectively form instantaneously.  What difference does it make how long they burn H after that?} 

 Formation of ultracompact galaxies at high $z$ necessarily occurs in a dense protogalactic environment, where the high density favors massive star formation. Compaction offers another pathway to forming so-called blue nuggets \citep{lapiner23}, but will inevitably be too rare and fine-tuned a process for the very large population of ``little red dots,'' the ultracompact galaxies at  $z\gtsim 5$.  So-called red nugget galaxies observed at low redshift are also compact and  some have obese black holes \citep{cohn23}, that is, black holes that are overly massive for the $M_{BH} - \sigma_\ast$ scaling relation.
% and may well be local relics of the high-$z$ population  of ultracompact galaxies. 
 Indeed, similarly obese black holes, with a much larger scatter relative to their  host galaxies than in the nearby universe, have been found at $z=4-6$ \citep{stone23}. We argue below that this evidence points to a scenario in which the formation of SMBHs, and indeed AGN, may even precede the formation of the dominant stellar components of galaxies. 

In theoretical support of this picture, we note that at  high $z$, bremsstrahlung cooling is very efficient within the central hundreds of parsecs in ultracompact galaxies.  At low $z$, especially at reduced gas fraction in gas-rich AGN hosts of ``typical'' size,  bremsstrahlung cooling is inefficient. This provides the motivation for our bimodal scenario of radiative turbulence and  momentum-conserving  outflows at high $z$ and energy-conserving outflows at low $z$.  
 
 We acknowledge, however, that observational support remains ambiguous, due in part to degeneracies among the diagnostics used to measure  
 high-redshift stellar masses, AGN luminosity, IMF, burstiness and reddening. Selection biases need to be more thoroughly understood before our inferences can be regarded as robust \citep{li2022}.

\subsection{Gas-rich initial conditions}
 
Our model simplifies the observational situation by making the premise that the little red dots are 10 times smaller, 10 times more gas-rich, but with $\sim 10\% $ of the mass of their low-$z$ counterparts that host SMBH.  The body of current data generally supports this view but is far from conclusive.  Most uncertain is our  claim that such objects are, at least initially, gas-rich. We infer this from the measured extreme stellar densities. However, whether gas is in a compact bulge-like distribution or in a more extensive disk is unclear. The observations nevertheless reinforce our confidence in a significant cooling transition, from early strong cooling to late-epoch cooling suppression.  
%This comparison echos pioneering arguments about galaxy formation \cite{ostriker, silka} that compared gravity-driven turbulent time-scales with global time-scales.
Although complications of enrichment and molecular gas cooling ultimately need to be taken into account, it is unlikely that our general conclusions will change.  In summary, for the observed and inferred column densities of $z\gtsim 5$ galaxies, cooling occurs rapidly and over very short length-scales for the dominant gas-rich and AGN-containing galaxy population.
 
Cooling of the outflows is generally inefficient at lower redshifts, where one  might expect to observe generic energy-conserving flows \citep{wagner12, costa14}.  These can reach large radii \citep{rupke17}, depleting the gas reservoir and quenching the star formation rate. Sporadic   
momentum-conserving episodes could still occur,  
triggering rarer cases of star formation in dense gas clumps.

Further evidence for the preponderance of gas rather than stars in massive  galactic halos at high $z$ can be inferred from the SMBH scaling relations. Under cooling-dominated conditions, SMBH scaling relations are embedded early and close-in.  \cite{maiolino23} finds a ratio of SMBH mass to stellar mass, {$M_{BH}/M_\ast$}, enhanced  by a large factor $\gtsim 100$ whereas the ratio of SMBH to dynamical mass, reflected in the $M_{BH} - \sigma_\ast$ correlation, follows the standard low redshift behavior.  This is consistent with SMBH growth at close to the Eddington rate, tightly coupled to feedback stirring up the circumnuclear gas and promoting nuclear star formation.  Here star formation is triggered by the compression of dense nuclear clouds, a phenomenon we may refer to as positive feedback.  In the low redshift universe, cases of positive feedback  are rare.  Other samples also show undermassive  SMBH at high $z$ in the mass range $M_{BH} =10^6-10^8 \rm M_\odot$ relative to the  local $M_{BH} - \sigma_\ast$ correlation, cf. \cite{harikane2023}, but nevertheless in general agreement with the preceding result when the large gas mass is taken into account.  Note, however, that other efforts to determine similar scalings at high redshift  do differ, with indications of overmassive SMBH at high $z$ from dynamical constraints, cf.  \cite{ubler2023}
and from a recent Chandra x-ray detection at $z=10.1$ \citep{priya23}. 
On balance, we cautiously argue  from the evidence that the most massive halos at high redshift are plausibly star-poor, guaranteeing a gas reservoir for triggering star formation at early epochs.

\subsection{Positive feedback at high redshift}

Observational signatures of positive feedback  include spatially extended star formation near radio jets \citep{duggal23}; star formation triggered at the edges of radio jet-induced bubbles \citep{venturi23}; interactions of low-powered trapped jets with the interstellar medium \citep{girdhar22};  enhanced nuclear star formation in quasar hosts \citep{molina23}; nuclear rings of enhanced star formation \citep{zhang23,pak23}, created by the outflowing nuclear jet driving through and compressing the interstellar gas \citep{dugan17}; and enhanced star formation rates in galactic outflows \citep{gallagher19}.  More generally, positive feedback can be monitored by emission line signatures in star-forming regions,  enhanced nuclear star formation accompanying accelerated SMBH growth, as well as associated supernovae and chemical enrichment.  

There are indications that the number of luminous galaxies at $z\gtsim 10$ exceeds those of standard simulation predictions \citep{harikane23}.  We note that the lensed sample of UV luminous galaxies at $z\gtsim 9$ studied by \cite{Chem23} included a high fraction of AGN \citep{Fujimoto23}.  

We speculate that positive feedback is a plausible pathway to account for central bursts of star formation that generate the early high luminosities, 
especially with inevitably bursty star formation \citep{sun23}. Rival suggestions seem more contrived, including feedback-free star formation, which requires several metal-free cycles  \citep{dekel23}, or the fine-tuning of a top-heavy IMF \citep{finkelstein23, yung24}.  While such other interpretations abound, we argue that the  observed  coevolution of AGN and star formation merits a more fundamental explanation.

\subsection{Transition}
The transition redshift from radiative to energy-conserving feedback is hard to estimate a priori, and likely depends on a number of uncertain factors. As discussed in Sec.~3.4, in early radiative stages the feedback energy may be mixed thoroughly with the ISM, leading to virial turbulence that pervades the galaxy.  As the transition approaches, this mixing becomes less complete and the situation begins to resemble the large-scale, cooled (i.e., momentum-conserving) shocks considered by \cite{fabian99}. Substantial mass loading in the outflow-driven shock, whether disk wind-driven \citep{raouf2023} or radio jet-driven \citep{venturi2023}, sets favorable  conditions for post-shock cooling.  The cooling column can be accumulated by some combination of shock compression and turbulent mixing \citep{begelmanfabian1990} facilitated by Rayleigh-Taylor and Kelvin-Helmholtz instabilities.  Because of large density contrasts between the AGN wind or jet and the dense ISM, the speeds governing the shock or mixing process are likely to be smaller than observed outflow speeds (typically a few thousand km s$^{-1}$) but may exceed the virial speeds used to estimate the cooling column in Sec.~2.  These considerations suggest that a typical cooled layer can have a column density several times larger than the estimate of Sec.~2, e.g., $N_{\rm cool} \sim 10^{23}$ cm$^{-2}$.  

The estimate above is for a single cooling layer. However, the {\it total} cooled column may be substantially larger than this due to the turbulent nature of the mixing process, which can create a two-phase medium characterized by a multiplicity of dense cooled layers interspersed with hot gas throughout the galaxy. If there are ${\cal N}$ such layers along a typical radius, then $N_{\rm cool}$ will increase by this factor. 
  
The transition to energy-conserving feedback occurs when the actual column density in the galaxy drops below $N_{\rm cool}$. Empirical data on interstellar medium column densities in massive and AGN host galaxies suggest that the gas column evolves with redshift approximately as $(1+z)^{3.3}$ \citep{gilli22}, and is some 100 times larger than local values at $z= 4-6. $ The estimated normalization is
$N_H= 10^{21} (1+z)^{3.3}\rm cm^{-2}$.  Then, if ${\cal N} \sim O(10)$, the total cooled column is $\sim 10^{24}\rm cm^{-2}$.  With this order-of-magnitude  estimate, we infer a transition redshift of $\sim 6$, consistent with observational constraints.   

This estimated transition epoch separates momentum-conserving flows at high $z$ and energy-conserving flows at low $z$.  Beyond $z\sim 6$, Compton thick values may be reached as in \cite{parlanti23}, explaining the likely absence of any X-ray detections of these AGN at such high redshift. Radio jet detections are predicted, but would be challenging, as discussed below.

\subsection{Global considerations}
The evolutionary scenario discussed above can be understood in terms of the evolving thermal structure of the galactic gas. %is more complicated. Star formation should remain  globally efficient via  momentum-driving. This because  the gravitational timescale, typically the rotation period, that controls e.g.~disk star formation, is now replaced by the jet/outflow crossing time, an order-of-magnitude shorter \citep{silk09}. AGN activity indeed correlates with amplified nuclear star formation \citep{molina23}. 
At early times, the cooling timescale of gas at the mean density is so short that the medium likely assumes a multi-phase structure, dominated by a ``hot" phase at roughly the virial temperature, filling most of the volume, and a cold, cloudy phase containing most of the mass.  
% MB The cloudy structure of a realistic interstellar medium adds further complexity.
%In a steady state there would presumably be a two-phase medium, with hot gas at $\sim T_{vir} $  occupying most of the volume but cold clouds containing most of the mass.  
Under these highly inhomogeneous conditions, we argue that most of the feedback energy is trapped and does not escape as a large-scale outflow, but rather drives turbulence and localized shocks throughout the galaxy.  Energy is lost both through cooling of the hot phase and strong radiative shocks due to collisions between cold clouds. 

A key controlling parameter is the ``covering factor" $C$ of the clouds, which could be large.  Physically, $C$ can be interpreted as the number of shocks encountered by a given parcel of cold gas per dynamical time.  The injection of energy due to feedback regulates $C$ by shredding the clouds until $C$ is large enough to dissipate the feedback energy.  A toy model illustrating these ideas is presented in the Appendix.  When $C \gg 1$, the feedback energy is radiated away and this corresponds to the radiative regime. As the galaxy grows and gas is used up to form stars, $C$ gradually decreases.  When $C \gtsim 1$, the galaxy approaches the transition discussed in Sec.~3.3, where cooling is still important but the feedback   drives a momentum-conserving outflow.  Finally, when $C\ltsim 1$, the feedback energy can sweep out the hot phase and entrain the cold gas in a wind, corresponding to the energy-conserving phase.

Generally, we would associate the radiative and momentum-conserving phases with enhanced star formation and the energy-conserving phase with negative feedback.  But the multi-phase character of the medium can also lead to negative feedback early in the radiative phase.  When $C$ is very large, the clouds can be shredded to the point where the size of a typical cloud is smaller than its Jeans length, so that star formation is suppressed (negative feedback).  This three-stage evolution   %in which star formation is initially suppressed, then progresses rapidly while $C \ltsim 1$, and is ultimately quenched by lack of gas during the energy-conserving stage. According to  the toy model presented in the Appendix, the transition from large to small $C$ is driven by increasing velocity dispersion and decreasing gas fraction as the galaxy builds in mass. }
fits naturally with models in which an overmassive BH forms first and a galaxy builds around it, with $\sigma_\ast$ increasing with time.  At first, star formation is inefficient (negative feedback due to clouds smaller than Jeans length) but then switches to rapid star formation as $C$ decreases (positive feedback), depleting the gas fraction and leading to a second phase of negative feedback: when $C$ becomes smaller than some threshold value, an energy-conserving wind sweeps through the galaxy.  If the galaxy assembly timescale is longer than the growth time of the black hole (a few tens of Myr for Eddington-limited accretion), then the SMBH regulates the size of the galaxy during early radiative stages: for a given SMBH mass the velocity dispersion increases until it hits the $M-\sigma_\ast$ line,  $M_{BH}\sim \sigma_\ast^5.$ 

\subsection{Negative feedback at low redshift}

Once AGN triggering of star formation fades, disk formation and instabilities are expected to limit star formation and to slow down any final collapse. Large-scale gravito-turbulence allows convergence to a global Schmidt--Kennicutt (SK) law \citep{nusser22}, regulates the star formation efficiency, and controls star formation rates in the nearby universe. 

The next phase is an energy-conserving outflow.  
Energy-driven feedback takes over at late times as the gas reservoir expands due to the feedback.  Star formation efficiencies, as deduced from the correlation of integrated star formation with molecular gas masses from NOEMA observations, favor gas depletion times of  0.1-1 Gyr up to $z\sim 6$, typically preferring mild starbursts \citep{berta23}.  JWST data confirm the presence of AGN-driven neutral outflows in massive galaxies at $z\sim 2$ \citep{davies23}, where star formation rates nevertheless follow a SK law \citep{schulze2019}. 
%\AN{I am not sure about this. To be honest, I don't think there is a case for energy driven outflows on a kpc scale. Sure, there is an initial phase of energy-conserving wind, even at high redshift (initially the wind is energy conserving, with cooling  dominating at $r>Rc$) . I might be wrong, but the  Tombesi et al  refers to an outflow at 300 pc  close to $R_c $. My point is that I do not see a difference between high z and low z if  the densities within 300pc are similar (as mentioned in the intro). I think there is a need to elaborate here.} 
 
Observations of outflows driven by AGN accretion disk winds at $\sim 0.1 c$  tend to support energy conservation \citep{tombesi15} from sub-pc to kpc scales and beyond, where molecular outflows are observed. The more recent data on outflows has a large dispersion, but generally supports black hole and galaxy coevolution \citep{capelo23}.

\subsection{Scaling laws}

In the momentum-conserving limit, the central black hole grows by accretion until inflow is halted and overpowered  by dynamical feedback.  The black hole mass growth saturates at  $M_{BH}=
{{f_{\rm gas}\sigma_T } \over {\pi G^2 m_p}  } \sigma_\ast^4,$ with reasonable assumptions about the wind speed and efficiency \citep{fabian99}. This expression approximately fits the observed  $M_{BH}-\sigma_\ast$ correlation in slope and normalization, according to theory \citep{costa14} and observations for  elliptical galaxies (and classical bulges) \citep{kormendy20} and AGN with $M_{BH} =10^7-10^9 M_\odot$ \citep{bennert21}.
 
In the energy-conserving limit, the feedback generates a late-time scaling law $M_{BH}={11f_{\rm gas}\sigma_T\over{\epsilon\pi G^2 m_p c}}\sigma_\ast^5$ \citep{silk98,costa14}, where $\epsilon$ is the radiative efficiency of the outflow. 
%\AN{Costa et al is a theory paper. Need to cite obs papers on this. Also, we need to be accurate with this. https://arxiv.org/pdf/1112.1078.pdf give a power  of 4.53 for ellipticals and SO and a power of 5.12 for their whole sample. Earlier papers by Tremaine at al got a think closer to 4. Personally, I think the derived power is plagued with observational selections. }.
%Other observational clues include the  SMBH mass scaling relations with stellar and halo masses, 
%AGN-correlated stochasticity  \mitch{Do we really claim evidence of "stochasticity" rather than just some time-averaged correlation?  Is there evidence to back this up?} of star formation rates including  rare episodes of positive feedback at low redshift as well as negative feedback at high redshift.
Late contributions to SMBH growth in this regime suggest that the final scaling law should be intermediate between  $M_{BH}\propto \sigma_\ast^4$ and  $M_{BH}\propto\sigma_\ast^5$.  Local observations indeed prefer $\alpha\approx 4.4$ \citep{kormendy}.  Our model actually predicts a steepening of the scaling law with redshift from $z\gtsim 5$ to $z\ltsim 5$. The steeper slope and lower normalization for the energy-driven limit, dictated by the factor $\sim \sigma_\ast / c$, 
suggest that SMBHs forming at low $z$ will be slightly undermassive  \citep{baldassare20}.  Outflows control late epoch quenching of star formation.  They might also be relevant for dwarf galaxies, where outflows are expected to be even more dominant at early gas-rich epochs because of the shallow potential wells.  This contrasts with  the obese (overmassive) mode of the $M_{BH}-\sigma_\ast$ correlation, inferred dynamically at high $z$, that we argue coexists with stimulated star formation and stellar mass growth.

\subsection{Indications from simulations}

Early studies showed that AGN outflows can produce positive feedback on star formation by overpressurizing circumgalactic gas \citep{bieri16}. 
Radio jets are  recognized to suppress star formation by stirring up ISM turbulence as well as producing both positive and negative feedback on star formation by driving  compression of ambient clouds \citep{mandal21}.  
%Recent simulations are inconclusive about the role of positive feedback in the cores of forming galaxies, although the consensus is that 
At later epochs ($z\ltsim 5$), the consensus view is that global negative feedback drives massive outflows, exhausts the gas supply,  quenches star formation and terminates galaxy growth.

Positive feedback involving strong compression of off-center gas clouds by quasar winds is indeed found in the most recent simulations \citep{mercedes23}, echoing the preceding analytic discussion and early 1-D simulations \citep{silk10}.

%Positive feedback from  BH-triggered central outflows could provide the transition to  early vigorous star formation.  This stimulus provides a  natural means of accounting for the apparently high stellar luminosities seen in some high redshift galaxies, that are not predicted in the usual numerical simulation suites. We predict that these star-forming circumnuclear clouds are overpressured by a momentum-driven jet or wind, triggering them to collapse.  

\subsection{Observational probes}

Early chemical enrichment provides a new dimension to our model. A  remarkable [N/O] $\gtsim +0.6$ enrichment \citep{cameron23} is seen in GN-z11 \citep{bunker23}, with one of the  highest confirmed spectroscopic redshifts for any galaxy.  One possible source  could be very massive stars with initial masses of hundreds of solar masses \citep{vink23}, associated with a top-heavy IMF \citep{bekki23}.  An alternative could be  supermassive stars \citep{marques23}, with initial masses of thousands of solar masses, or direct-collapse black-hole precursors with masses possibly reaching $10^5-10^6 M_\odot$.  Either option requires rapid early star formation in massive central star clusters \citep{belokurov23}, possibly in intermittent episodes \citep{kobayashi23}.   Similar episodes are predicted in our boosted nuclear star formation model.   
 
Such N excesses are common to quasar emission line regions \citep{dietrich03}, further justifying the AGN-star formation connection.  There is one more consequence of the positive feedback episodes driven by radiative feedback. In addition to triggering star formation, these can also lead to compactification of gas clouds whose mergers  provide sources of clumpy accretion in the nuclear environment. These may boost BH growth via early super-Eddington accretion in massive halos \citep{bennett23}.

\section {Massive black hole seeds}

The large population of ultracompact, dust-reddened red galaxies sets the scene for the preceding discussion of  coevolution of star formation and AGN outflows.  We have argued that this implies the existence of a large population of massive black holes at very early times.  How did they form?  We first focus on Population III.  Massive seeds are one possible precursor to the observed high-$z$ massive BH population. Suppression of H$_2$ by UV in the Lyman-Werner bands is one ingredient that can promote the formation of massive seeds.  The near-dominance (and possible predominance) of AGN at the earliest epochs suggests that massive BH seeds could coevolve with Population III if the latter provide a sufficient local UV background at $z=10-15$.  A plausible environment would be that of early-forming protoclusters which contain AGN that form coevally with the first galaxies. 

Direct collapse BH seeds can also be abetted at early times by  magnetic pressure in the central gas disks that coexist with the first AGN. This could result from early dynamo activity, with analytic studies favoring magnetic levitation suppressing disk fragmentation to stars \citep{begelman17} and catalyzing  BH growth  by reducing the density and speeding up the central inflow \citep{begelman23}.   

Magnetic boosting and/or turbulence driven by low angular momentum cold flows provide an alternative channel for  suppressing premature fragmentation into stars.  Recent MHD simulations suggest that these may generate strong toroidal fields that dominate the inner accretion disk and thereby suppress disk fragmentation \citep{hopkins23}. 

The early formation of a large population of (ultracompact) galaxies is problematic in the standard $\Lambda$CDM model.  A possible resolution could involve excess small-scale power in the primordial density fluctuation power spectrum on dwarf galaxy scales, as has been explored in earlier studies of ways of boosting early formation of galaxies \citep{chevallard} and AGN \citep{habouzit} by primordial galaxy-scale nongaussianities \citep{sabti} or by primordial spectral bumps \citep{tkachev24}.  It is well known that the formation of SMBH, especially at high $z$, cannot be explained if accretion growth from stellar-mass black holes is Eddington accretion-limited. 
 
This suggests that any solution should simultaneously resolve the issues of AGN host compactness at high $z$ and the early formation of SMBHs.  We consider several possible pathways:
\begin{itemize}  
\item 
The seed black holes may be generated at sufficiently early epochs by supermassive stars \citep{begelman06,begelman10} forming from excess small-scale power in the distribution of primordial density fluctuations \citep{tkachev23}, or by scale-dependent nongaussianity  \citep{habouzit16}.  The supermassive star ansatz might provide one way to account for the N excess in GN-z11 \citep {nagele23}, but production of a large number of such seeds would be constrained by early heavy element production.
\item Alternatively, there could be a seeding population of primordial black holes which need only contribute $\sim10^{-4}$ of the dark matter, specified by the seed abundance requirement and the observationally constrained $10^3-10^6\rm M_\odot$ mass range, and restricted by microlensing, dynamical friction and accretion limits \citep{yuan2023}.  The transformation from primordial black holes to ultracompact galaxies is accounted for by gas accretion with predicted  growth since last scattering by a factor $\sim 100$ 
\citep{carr18}.  The ultracompact galaxies are expected to contain obese central back holes, relative to the usual dynamical scaling relation,
%{\color{red}\bf This makes for an interesting prediction as to the degree of obesity, that is an offset from the SMBH dynamical scaling relation, namely the $M_{BH}, \sigma $ correlation, 
with the ratio of black hole to galaxy mass decreasing towards lower redshift.  

\item  We might also consider the super-Eddington growth of moderate mass seeds inside quasi-stars \citep{begelman08}.  These would resemble red giants (or Thorne-\. Zytkow objects) but powered by BH accretion with gaseous envelopes that are orders of magnitude more massive than the central BH.  Accretion at the Eddington limit for the envelope mass can rapidly produce $\sim 10^5-10^6$ BH seeds. 

\end{itemize}
 
%{The challenge in massive seed formation is  to avoid premature fragmentation into stars. } This is  central to the fundamentals  of SMBH formation. Early suppression of fragmentation is conjectured to occur via destruction of the predominant early coolant $H_2$  by Lyman-Werner band UV radiation.  This occurs in Population III scenarios where there are neighboring sources of UV radiation in close proximity.%We have seen that  there are also  suppression scenarios involving magnetic turbulence that pressurizes   and even levitates self-gravitating disks and and thereby quenches star formation. These models require the early generation of protogalactic fields by dynamo mechanisms from cosmological initial conditions.

The primordial black hole scenario --- which has support in diverse inflationary scenarios --- is more direct, given the initial hypothesis that SMBHs ($\gtsim 10^6\rm M_\odot$) form before significant star formation. The required mass range of primordial intermediate-mass black holes (IMBHs: $10^3-10^6\rm M_\odot$) is allowed at a contribution of up to  $\sim 10^{-4} $ of the dark matter  density.  The observed (comoving) number density of high-$z$ AGN is $\sim 10^{-5}\; {\rm Mpc}^{-3}$, corresponding to a mass density $\sim 10 (M/10^6 {\rm M}_\odot ) M_\odot \; {\rm Mpc}^{-3} $, whereas the DM density is  $\sim 10^4 \;{\rm M}_\odot \; \rm Mpc^{-3}$.  
 
Early fragmentation might even be a positive  factor in the context of boosted small-scale primordial density fluctuations. For example,  early-forming exceptionally dense star clusters could be factories for runaway astrophysical black hole mergers and early formation of IMBH seeds that would be enhanced in mass by gas accretion and dwarf galaxy mergers. They  would produce distinct gravitational wave signals for future-generation experiments \citep{kritos23}.  Nearby dwarf galaxies often have nuclear star clusters \citep{Carlsten22} and IMBHs \citep{Reines2020}.

\section{Discussion}

We find three phases of early coevolution of SMBHs and their galaxy hosts:
 
 a)   {$z\gtsim 15$: Black hole growth and the rise of AGN }
   
b) {$5\ltsim z\ltsim 15$: Star formation bursts triggered by radiative turbulence and momentum-conserving AGN outflows into a dense clumpy interstellar medium}  
   
c) {$z \ltsim 5 $: Star formation quenched and gas depleted by energy-conserving AGN outflows}

The case for SMBHs forming simultaneously with galaxies seems compelling.  We focus here on the interplay between AGN activity and star formation, and argue for a bimodal synergy.  AGN feedback may have evolved  from a short-lived but vigorous phase of positive feedback via radiative turbulence and momentum-driven outflows in  ultracompact galaxy hosts. Radiative shocks, both from cloud collisions and outflows, triggered positive feedback and drove the first episodes of vigorous star formation at $z \gtsim 10$.  Energy-driven outflows in initially gas-rich galaxies then depleted halo gas reservoirs and stimulated negative feedback, transitioning at $z \ltsim 6$.  
 
Coevolution of AGN outflows and star formation may be the missing link to the high redshift Universe.  Nearby examples abound of both negative and positive feedback associated with AGN outflows.  Positive feedback may be a missing ingredient for early boosting of star formation that is intimately linked to the earliest AGN activity.  The rapid evolution of stellar mass, the compactness of the galaxy population  and the growth of SMBHs at early epochs are outstanding issues that remain to be  adequately addressed in  cosmological simulations \citep{byrne23}.

\subsection{Obscuration}

For massive galaxies, we have found that at high redshifts, cooling of QSO winds ($ \sim 3000$ km s$^{-1}$) occurs above $z\sim 6$ for host galaxy gas column densities $\ltsim 10^{23} \rm cm^{-2}$.   This range characterizes the transition to momentum-driven feedback as $z$ increases.  Such molecular column densities are inferred from theory and observations \citep{gilli22}, and indicate that it is not the ``classical" torus obscuring the AGN, but the  galaxy's ISM itself.  The minimum obscuration along the line of sight for transitioning from momentum-driven to energy-driven outflows is
 $$N_{\rm cool} \approx 10^{23}{\rm cm}^{-2}\left(v_s\over{3000 \, {\rm km \, s}^{-2}}\right)^2, $$ where $v_s$ is the feedback outflow velocity.  Even Compton-thick obscuring host galaxies with no X-ray detections may well exceed this cooling column by more than an order of magnitude.  All of these high redshift objects should have cooled, momentum-driven winds and outflows.  For the relevant range of cooling columns $N_{\rm cool} \approx 10^{23}-10^{24}\, \rm cm^{-2}$, with Compton depths 0.1 to 1.0, the AGN will be obscured.  While this simple estimate assumes a spherical outflow, geometrical arguments suggest that in the more general case of directed outflows, high-$z$ AGN are always buried or obscured along the cooling momentum-driven outflow. In this case, viewed from  the side, one would not see the full outflow velocity. Observable effects might, however, include turbulence, narrow line emission and dust.

\subsection{Observational predictions}

Simple radiative or momentum-driven feedback for massive high-redshift galaxies has such a high requirement on the column density that the central AGN driving the feedback is hidden and will not generate detectable emission lines or X-rays.  It is hidden by the host galaxy ISM, and not by a Type I/Type II AGN torus.  Our best bet to detect hidden momentum-driven AGN feedback may be via searching for very high specific star formation rates due to the cooling shock overpressure.  In principle, it should be much easier to find a positive feedback signature than has hitherto been the case in attempts to find direct evidence of the negative signature associated with quenching of star formation. The former effect is of order $10(v_s/3000 \, {\rm km \, s}^{-1})/  (v_{\rm cir} /300 \,{\rm km \, s}^{-1})$ faster, where $v_{\rm cir}$ is the galaxy potential-defined circular velocity of the massive host galaxy. 

Other observational predictions include redshift dependence in the SMBH mass scaling relations with stellar and halo masses, AGN-correlated star formation rates, and chemical abundance signatures.  Improved statistics, imaging and spectroscopic follow-up are needed to justify the systematic global interpretation that is being proposed here with admittedly simplistic but plausible arguments.  Finally, we note that specific star formation rates are not yet available for the JWST galaxies in the $z=5-15$ range.  These would be an important indicator of positive feedback, especially if correlated with the presence of an AGN or various multi-wavelength luminosities, including in the radio and/or X-ray bands, as well as optical/IR emission lines and infrared dust continuum.  The star formation and AGN duty cycles may be out of phase and dust-obscured, so detection of any such correlations would be challenging.  With current X-ray missions, there is little hope of detection unless there is an object in the center of an existing deep field such as the CDFS 7Msec deep field.  JWST might help with detection of faint broad wings of IR lines but the Compton-thick obscuration is a formidable obstacle.

One option currently being explored is the use of SKA radio observations to detect these deeply buried  high redshift QSOs (G. Mazzolari, private communication), but the expected detection rates are small. Finally we note that at lower redshifts than $\sim 6$, there is no relation in our model between AGN obscuration due to the ``torus" and our cooling-driven wind/outflow feedback mechanism.  The latter is dependent on the host galaxy ISM.  Only at the higher redshifts considered here can we relate  obscuration and ISM shock cooling. This regime  may eventually be revealed by deep ALMA observations. 

\noindent{\bf \em Acknowledgements}:
We thank Roberto Gilli, Luis Ho,  Erini Lambrides, John Silverman, Alex Wagner and Feng Yuan for inspiring comments and the referee for an encouraging report.  AN  is supported by the Israel Science Foundation (ISF) grant 893/22 and the Asher Space Research Institute.  RFGW acknowledges  support through the generosity of Eric and Wendy Schmidt, by recommendation of the Schmidt Futures program.

\begin{appendix}
\section{Toy model for radiative feedback}
\label{appendix}

In the radiative feedback limit, we assume a two-phase medium with cold clouds at $10^3 T_3$ K and a  diffuse hot phase at $T_{\rm vir} = 10^7 M_{10} r_{150}^{-1}$ K, where we use the same notation as in Sec.~2.  For simplicity, we assume that both phases absorb similar amounts of feedback power, roughly 10\% of the Eddington luminosity for a $10^8 m_8 M_\odot$ SMBH.  If the hot phase radiates away this energy via bremsstrahlung, its density is given by $n_h \approx 600 m_8^{1/2} M_{10}^{-1/4} r_{150}^{-5/4}$ cm$^{-3}$.  For characteristic parameters, we see that this is much smaller than the mean density $\bar n$ from Sec.~2.  In pressure equilibrium, the cold phase density is $n_c \approx 6\times 10^6 m_8^{1/2} M_{10}^{3/4} r_{150}^{-9/4} T_3^{-1}$ cm$^{-3}$ and the volume filling factor is small, $f_c = f_{\rm gas} \bar n /n_c \approx 0.01 f_{\rm gas} m_8^{-1/2} M_{10}^{1/4} r_{150}^{-3/4} T_3$. 

We now introduce the cloud covering factor, $C$, discussed in Sec.~3.4, which is, in effect, an optical depth for cloud--cloud collisions.  If we regard the clouds as spheres of radius $r_c$, we have $r_c \approx C^{-1} f_c r \approx C^{-1} f_{\rm gas} m_8^{-1/2} M_{10}^{1/4} r_{150}^{1/4} T_3 $ pc. The Jeans length for these clouds is given by $r_J \approx 0.1  m_8^{-1/4} M_{10}^{-3/8} r_{150}^{9/8} T_3$ pc.  We thus find that $r_c < r_J$ when $C > 10 f_{\rm gas} m_8^{3/4} M_{10}^{5/8} r_{150}^{-7/8}$, implying that the clouds are too small to gravitationally fragment and form stars when $C$ is sufficiently large. 

We propose that the clouds are stirred up to virial random velocities by the feedback, then radiate away this energy at a rate $\sim C f_{\rm gas} \sigma_\ast^5/G $ via cloud--cloud collisions. Equating this dissipation rate to the feedback power supply determines $C$. Adopting our earlier assumption about the magnitude of the feedback power supply, we can also express this relation as $m_8 \sim 0.4 C f_{\rm gas} (\sigma_\ast / 200 \, {\rm km \, s}^{-1})^5$, which is very close to the modern-day $M-\sigma_\ast$ relation when $C \sim$ a few.  This relation also suggests that star formation can be suppressed by high values of $C$ at early times, if SMBH seeds are initially overmassive with respect to the $\sigma_\ast$ of the protogalaxies forming around them.   

\end{appendix}

\bibliography{SMBHgal}{}
\bibliographystyle{aasjournal}

\end{document}